\documentclass[useAMS,usenatbib]{mn2e}
\usepackage{amsmath}
\usepackage{times}
\usepackage{graphicx}
\usepackage[colorlinks=true, linkcolor=blue, citecolor=blue, urlcolor=blue]{hyperref}
\title[Radiative transfer modeling of the Sombrero galaxy]{Panchromatic radiative transfer modeling of stars and dust in the Sombrero galaxy}

\author[I. De Looze et al.]{
Ilse De Looze, Maarten Baes, Jacopo Fritz and Joris Verstappen \\
Sterrenkundig Observatorium, Universiteit Gent, Krijgslaan 281 S9, B-9000 Gent, Belgium}

\begin{document}

\date{\today}

\pagerange{\pageref{firstpage}--\pageref{lastpage}} \pubyear{2011}

\maketitle

\label{firstpage}

\begin{abstract} 
  We present a detailed study of the dust energy balance in the Sombrero galaxy M104. From a full radiative transfer analysis, including scattering, absorption and thermal re-emission, we construct models that can reproduce images at optical/near-infrared wavelengths, the observed stellar SED and the minor axis extinction profiles in the $V$ and $R_{C}$ band. A standard model, that contains only an old stellar population to heat the dust, underestimates the observations of dust emission at infrared wavelengths by a factor of $\sim$ 3.  Supplementing this basic model with a young stellar component of low star formation activity in both the inner disk (SFR $\sim$ 0.21 M$_{\odot}$ yr$^{-1}$) and dust ring (SFR $\sim$ 0.05 M$_{\odot}$ yr$^{-1}$), we are capable of solving the discrepancy in the dust energy budget of the Sombrero galaxy at wavelengths shortwards of 100 $\mu$m. To account for the increased FIR/submm emission beyond 100 $\mu$m, we propose a additional dust component distributed in quiescent clumps.  This model with a clumpy dust structure predicts three-quarters of the total dust content ($\sim$ 2.8 $\times$ 10$^{7}$ M$_{\odot}$) to reside in compact dust clouds with no associated embedded sources. Although the assumption of a clumpy dust structure in the Sombrero galaxy is supported by high-resolution optical data, we cannot rule out the possibility that dust grains with a higher dust emissivity account for part of the discrepancy in the energy budget at submm wavelengths.  \end{abstract}


\section{Introduction}

Although the dust content of a galaxy is often negligible compared to the other components in the ISM, dust grains play a prominent role in several astrophysical processes in a galaxy.  On the one hand, they regulate the cooling of gas in the ISM through photoelectric heating and inelastic interactions with gas particles. This process of cooling allows the gas in a galaxy to eventually condense into stars.  On the other hand, the extinction of stellar photons by dust particles distorts our view on the stars populating a galaxy.  A profound knowledge of the dust properties is thus of great importance to understand galaxy evolution processes such as star formation and determine the specific amount of stellar photons reprocessed by dust particles.

Optical studies reveal the extinguishing effect of dust grains on stellar light at different wavelengths. Radiative transfer codes have been used to model the dust distribution in spiral galaxies from optical data \citep{1987ApJ...317..637K, Xilouris1997, Xilouris1998, Xilouris1999, Alton, Bianchi2007}. These optical dust studies all show that the scale height of the dust disk is about half of the scale height of the stellar disk, while the scale length of this dust disk is often more extended than that of the stellar distribution. These thinner dust disks are optically thin (central optical depth $<$ 1) at optical wavelengths up to an inclination of about 60$\degr$.

An alternative way of studying the dust characteristics in a galaxy is to probe the dust emission at FIR/submm wavelengths.  With the recent advent of space (Herschel; \citealt{2010A&A...518L...1P}) and ground based (ALMA) facilities in this particular wavelength domain, the dust in nearby as well as distant galaxies can be traced with a significantly improved sensitivity and spatial resolution.  Tracing the reprocessed stellar light at FIR/submm wavelengths, one would expect that the thermal emission of dust exactly compensates for the amount of photon energy absorbed at UV/optical wavelengths.

Ideally, a complementary study of the dust component from optical/UV images and its thermal re-emission at FIR/submm wavelengths allows to put strong constraints on the properties of dust in a galaxy.  Up to now, the energy budget has already been probed in several edge-on spirals \citep{Popescu, 2011A&A...527A.109P, 2001A&A...372..775M, 2002MNRAS.335L..41P, Alton, Dasyra, 2010A&A...518L..39B}. The edge-on view of these objects offers the opportunity to trace dust out to large radii and, hereby, make an accurate estimate of the scale length and height of the dust, which can hardly be distinguished in the face-on case.  Unfortunately, the high inclination of the disk also implies that most of the details of the spiral arms in the stellar disk are smeared out and most details of the dust distribution vanish when integrating along the line-of-sight.

In this paper, we present a detailed radiative transfer analysis of the Sombrero galaxy (M104).  At a distance of only 9.2 Mpc (which is the average of measurements from \citealt{1996ApJ...458..455F} and \citealt{1997AJ....114..626A}), it is one of the most nearby early-type spirals. The Sombrero galaxy has been studied in detail, with a particular focus on its {supermassive black hole} (SMBH) \citep{1996ApJ...473L..91K}, rich globular cluster population \citep[e.g.,][]{2008MNRAS.385..361S,2009AJ....138..758C,2010MNRAS.401.1965H,2010ApJ...720..516B}, stellar population \citep[e.g.,][]{1986AJ.....91..777B, 1996ApJ...458..455F, 2010ApJ...722..721M} and discrete X-ray sources \citep[e.g.,][]{2007MNRAS.376..960L, 2010ApJ...721.1368L}. Its proximity implies that also the distribution and characteristics of dust can be examined in much detail. Its inclination of $\sim$ 84$^\circ$ combines the advantages of these edge-on spirals with an insight of the inner regions. This allows to put strong constraints on the distribution of stars and dust in the disk.  While IRAS observations had to settle for a global SED study in the infrared/submm waveband due to {its limited angular} resolving power \citep{1988ApJS...68...91R, 1997AJ....114..592S}, \citet{Bendo} could separate the emission from different components contributing to the dust emission at infrared wavelengths. From their image modeling procedure on the highest resolution {Spitzer/MIPS} data (24 and 70 $\mu$m), they could separate the emission from the nucleus, the bulge, the inner disk and the dust ring contributing to the different infrared wavebands.  In the submillimeter band, \citet{Vlahakis} could disentangle the AGN from the dust ring component in the LABOCA 870 $\mu$m and MAMBO-2 1.2 mm images.  This wealth of available FIR observations (IRAS: \citealt{1988ApJS...68...91R}; MIPS: \citealt{Bendo}; LABOCA \& MAMBO: \citealt{Vlahakis}) and the very regular and symmetric {shape of the dust lane} in M104 permit a spatially resolved analysis of the dust energy balance in this nearby galaxy.

In the past, the Sombrero dust ring has been probed in many related studies. \citet{1989A&A...209....8M} derived a radial distribution of dust in the disk from observations of the polarization along the major axis. \citet{1991A&A...241...42K} were the first to study the extinction law in M104. They found a good agreement with the Galactic extinction law, except for the \textit{B} and \textit{V} band, which required the assumption of embedded stars in the dust lane or additional foreground light. \citet{Emsellem} used high resolution HRCAM $B$, $V$, $R_{\text{C}}$ and $I_{\text{C}}$ data to construct a model for both the stellar and dust component in the Sombrero ring.  From discrepancies between his model and observed extinction profiles, he concluded that scattering is an important extinction process and neglecting its effect in highly-inclined dust disks may cause a substantial underestimation of the true dust content in such a galaxy.

In this paper, we will extend the analysis in \citet{Emsellem} to a full radiative transfer study, including absorption, scattering and thermal dust re-emission. Continuing on the image modeling at infrared wavelengths from \citet{Bendo}, this multi-wavelength analysis aims to model the characteristics and distribution of dust in M104 accounting for both the extinction and dust emission properties. We use the SKIRT radiative transfer code \citep{2003MNRAS.343.1081B, Baes2011} to construct a 3D model for the stellar and dust component in the Sombrero galaxy. Subsequently, we compare the FIR/submm emission predicted by our radiative transfer model to the observed FIR/submm images and fluxes. This allows an unprecedented analysis of the dust energy balance in the Sombrero galaxy.

The strategy for constructing the best fitting model to the optical/NIR data is explained in {\S}2.  In $\S$3, we present the results from our radiative transfer modeling and discuss the dust energy balance for the various models.  $\S$4 gives an overview of previous results from energy balance studies in edge-on spirals and makes a comparison with the main conclusions for the energy budget in the Sombrero galaxy. Our final conclusions are summarized in $\S$5.

\section{Modeling procedure}

\subsection{Data}

In order to recover a reliable representation of the distribution and properties of both stars and dust in the Sombrero ring, we will fit a 3D model to the stellar SED, the images and minor axis extinction profiles in both the $V$ and $R_{C}$ band.  The stellar emission in the SED is constrained by the GALEX $FUV$ and $NUV$, optical $BVRI$, 2MASS $JHK$ and IRAC 3.6 and 4.5 $\mu$m fluxes, which were taken from the \citet{Dale} {broadband spectral atlas} (see Table 1 for an overview).  IRAC 5.8 and 8.0 $\mu$m measurements are also available, but were not used in the fitting procedure since the emission at those wavelengths is not only stellar in origin. Also the GALEX $FUV$ and $NUV$ data were omitted from the fitting procedure, since they critically depend on the amount of extinction and, therefore, the specific amount of young stars in a galaxy (see Section \ref{SFModel.sec} later on).  The $V$ band images were obtained from the archive of the Spitzer Infrared Nearby Galaxies Survey \citep{2003PASP..115..928K}, while the Spitzer 24 and 160 $\mu$m images could be retrieved from the Spitzer Local Volume Legacy survey \citep{2009ApJ...703..517D}.

\begin{table}
\caption{Observed fluxes for the Sombrero galaxy}
\label{dataptn}
\begin{center}
\begin{tabular}{@{}lccc}
\hline \hline \\
Filter & $\lambda$~($\mu$m) & {$F_{\nu}$}~(Jy) & ref\footnotemark[1]  \\
\hline
FUV & 0.15 & 0.0056 $\pm$ 0.0008 & 1 \\
NUV & 0.28 & 0.0177 $\pm$ 0.0025 & 1 \\
B & 0.45 & 2.25 $\pm$ 0.23 & 1\\ 
V & 0.55 & 2.76  $\pm$ 0.28 & 1\\
R & 0.66 & 3.41  $\pm$ 0.34 & 1 \\
I & 0.81 & 4.30  $\pm$ 0.43 & 1\\
J & 1.25 & 8.06  $\pm$ 0.81 & 1\\
H & 1.65 & 9.19  $\pm$ 0.927 & 1\\
K & 2.17 & 7.57  $\pm$ 0.76 & 1\\
IRAC 3.6 & 3.6 & 3.94 $\pm$ 0.53 & 1\\
IRAC 4.5 & 4.5 & 2.31 $\pm$ 0.32 & 1\\
IRAC 5.8 & 5.8 & 1.75 $\pm$ 0.22 & 1\\
IRAC 8.0 & 8.0 & 1.30 $\pm$ 0.16 & 1\\
IRAS 12 & 12 & 0.74 $\pm$ 0.19 & 2\\
MIPS 24 & 24 & 0.65 $\pm$ 0.07  & 3\\
IRAS 25 & 25 & 0.50 $\pm$ 0.13 & 2 \\
IRAS 60 & 60 & 4.26 $\pm$ 1.07 & 2 \\
MIPS 70 & 70 & 6.7 $\pm$ 1.3 &  3\\
IRAS 100 & 100 & 22.86 $\pm$ 5.72 & 2 \\
MIPS 160 & 160 &  35.1 $\pm$ 7.0 & 3 \\
LABOCA 870 & 870 & 0.924 $\pm$ 0.092 & 4 \\
MAMBO 1200 & 1200 & 0.442 $\pm$ 0.044 & 4 \\
\hline
\end{tabular}
\end{center}
\footnotemark[1]{References: \citealt{Dale} (1), \citealt{1988ApJS...68...91R} (2), \citealt{Bendo} (3), \citealt{Vlahakis} (4)}
\end{table}

\subsection{Radiative transfer: SKIRT}

SKIRT is a 3D Monte Carlo radiative transfer code, which was initially developed to investigate the effects of dust extinction on the photometry and kinematics of galaxies \citep{2002MNRAS.335..441B, 2003MNRAS.343.1081B}. Over the years, the radiative transfer code evolved into a flexible tool that can model the dust extinction, including both absorption and scattering, and the thermal re-emission of dust, under the assumption of local thermal equilibrium (LTE) \citep{2005AIPC..761...27B, 2005NewA...10..523B}. This LTE version of SKIRT is capable of modeling the dust properties in a {suite} of different environments: circumstellar disks \citep{2007BaltA..16..101V, Vidal2011}, clumpy tori around active galactic nuclei \citep{Stalevski2011} and a variety of galaxy types \citep{2010A&A...518L..39B,2010A&A...518L..54D,2010MNRAS.403.2053G, 2010A&A...518L..45G}.  Recently, the code was adapted to include the emission from very small grains and polycyclic aromatic hydrocarbon molecules \citep{Baes2011}.

\subsection{Model specifications}

\subsubsection{Stellar distribution}
\label{subsec2}

A model for the emission at optical wavelengths in M104 needs to account for the contribution from the bulge, disk, nucleus and the halo.  Considering that only minor departures from axial symmetry have been perceived along the disk and in the central region due to dust patches, {resembling dust lanes in a stellar bar \citep{2000A&A...357..111E}}, we assume an axi-symmetric geometry for the stellar component in the Sombrero galaxy.  We adopt the $15$ {Multi-Gaussian Expansion (MGE)} components, both in the $V$ and $R_{C}$ band, reported in \citet{Emsellem}.  Only the global intensity of the stellar component is scaled linearly until our models achieve the best fit with the stellar SED using a $\chi^2$ minimization procedure.

The intrinsic SED of the stars is parametrized by a \citet{1998MNRAS.300..872M, 2005MNRAS.362..799M} single stellar population with an age of 10 Gyr and a close to solar metallicity $Z=0.016$, based on the chemo-evolutionary analysis in \citet{Vazdekis}.  This value for the metallicity is somewhat higher than the estimate [Fe/H]~$\sim -0.5$ from \citet{2010ApJ...722..721M}, who determined the modal metallicity value for a field 6.8 effective radii from the center of M104.  Considering that this field is mainly populated by halo stars and, therefore, constitutes mainly of old metal-poor objects, we consider the metallicity estimate from \citet{2010ApJ...722..721M} within the same order of magnitude of the nearly solar metallicity ($Z\sim0.016$) adopted in this work.

Besides the main stellar body of the Sombrero galaxy, we also added {a nuclear source} to the model. At optical wavelengths, the nucleus of M104 was modeled as a point source in our radiative transfer model.  Also at IR/submm wavelengths, a significant portion of the emission arises from the nucleus.  Especially the origin for the high 850 $\mu$m compared to the 160 $\mu$m flux in the nucleus remains an unsettled issue \citep{Bendo, Vlahakis}.  In order to account for this peculiar emission from the nucleus at submm wavelengths, we also include the nucleus of the Sombrero galaxy as a point source emitting brightly in the IR.  The shape of the SED was obtained through an interpolation between the fluxes from the nucleus at IR/submm wavelengths \citep{Bendo, Vlahakis}.

\subsubsection{Dust distribution}
\label{subsec4}

Our dust model will need a representative parametrization of the dust distribution in both the ring and inner disk of M104.  Similar to the stellar component, the very regular and symmetric dust lane in the Sombrero galaxy justifies our choice of an axi-symmetric dust geometry for the dust ring.  Detailed analyses of the extinction profiles along the vertical axis showed that the dust in the dust lane is distributed in a ring with three peaks, which average out to a double ring structure along vertical profiles further from the galaxy's center \citep{1991A&A...241...42K, Emsellem}.

We model the dust ring as a linear combination of $4$ Gaussian functions: 
\begin{equation}
  \rho_{\text{d}}(R,z)
  =
  \sum^{4}_{i=1}~\rho_{0,i}~\exp 
  \left[-\frac{(R-R_{i})^{2}}{2 \sigma^{2}_{i}} \right],                   
  \quad
  -\frac{h_{z}}{2} \leq z \leq \frac{h_{z}}{2}.
\label{eqdens}
\end{equation}
For each Gaussian component, the radial distribution is Gaussian, while the density in the vertical direction is assumed to be constant. This constant density is justified by the small {extent} of the dust lane in vertical direction ($h_{z}$ $\sim$ 20 pc) \citep{Emsellem}. Three of the Gaussians are used to fit the peaks in the minor axis extinction profile, the fourth is used to describe the broader, ring-wide distribution of dust.

The spatial distribution of each individual Gaussian component is determined by three parameters: the radial distance $R_{i}$ at which the maximum value of the Gaussian function is situated, the width $\sigma_{i}$ of the Gaussian function and the relative mass contribution $\rho_{0,i}$. From the extinction curves, one can already determine fairly accurate estimates of the radial distance of the density peaks. Therefore, the only unknowns which are allowed to vary in our fitting procedure are the width $\sigma_{i}$ and central density $\rho_{0,i}$ of each Gaussian component.

Although the bulk of the dust content in the Sombrero galaxy is distributed in the symmetric ring, {the inner disk also} contains an amount of dust and young stars.  In correspondence to the model from \citet{Bendo}, we represent the inner dust disk as an exponential disk with a radial scale length of 2000 pc. The scale height is confined to 20 pc, following the height of the dust ring. The dust content of the inner disk was determined from a modified blackbody fit to the MIPS 70 and 160 $\mu$m fluxes and was set to 4.07 $\times$ 10$^{5}$ M$_{\odot}$. 

Throughout the ring and inner disk of the Sombrero galaxy, we assume a single dust mixture, consisting of different dust particles with a particular grain size distribution.  The abundances, extinction and emissivity of the dust mixture are taken from the \citet{2007ApJ...657..810D} model, which accurately reproduces the optical properties of the dust in our own Galaxy.

\subsection{Model fitting}
 
For a given set of parameters for the stellar and dust distribution, we run simulations with the radiative transfer code SKIRT. SKIRT self-consistently calculates the temperature distribution and thermal emissivity of the dust at every position in the model from the local interstellar radiation field (ISRF). The result of each simulation is the observed SED and a set of images at various wavelengths ranging from UV to mm wavebands.  In order to find the best fitting model, we have compared the model results with the observed minor axis extinction profiles and images in the $V$ and $R_{\text{C}}$ band and the stellar SED. Only the minor axis extinction profiles in the $V$ and $R_{\text{C}}$ band were used in the fitting procedure, because only in those wavebands the signal-to-noise was sufficient in the entire dust lane along the minor axis.  Allowing the several parameters to take values in a confined parameter space, we could narrow down the range of valid parameters and finally obtain the best fitting model that simultaneously reproduces both the stellar emission and dust extinction (through a least-square fitting procedure).

\section{Modeling results} 
\label{modeling}

\subsection{Standard model}
\label{StandardModel.sec}

In a first set of radiative transfer simulations, we only include the contribution from the old stellar population in the Sombrero galaxy.

\begin{figure} 
\centering 
\includegraphics[width=0.45\textwidth]{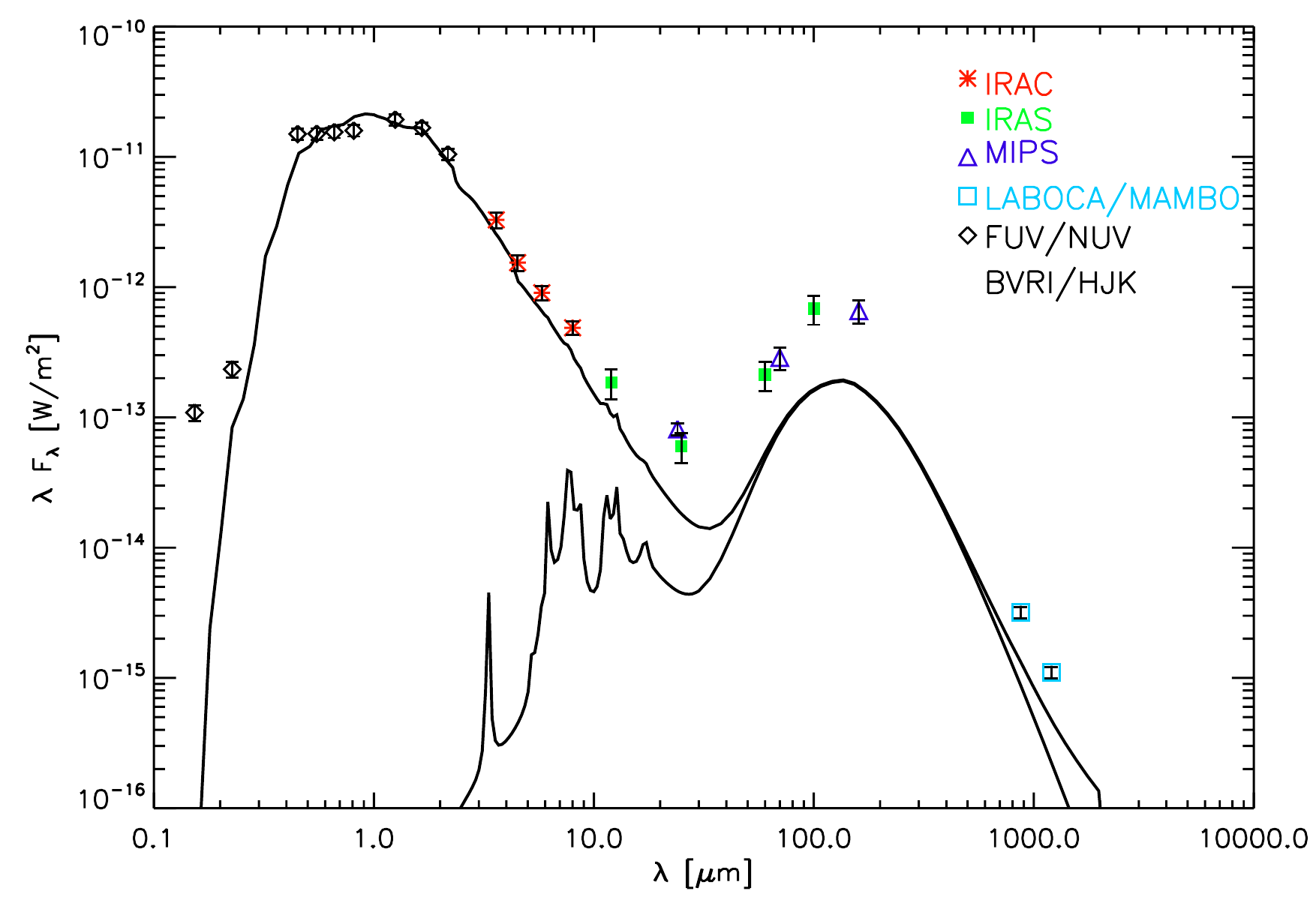} 
\includegraphics[width=0.42\textwidth]{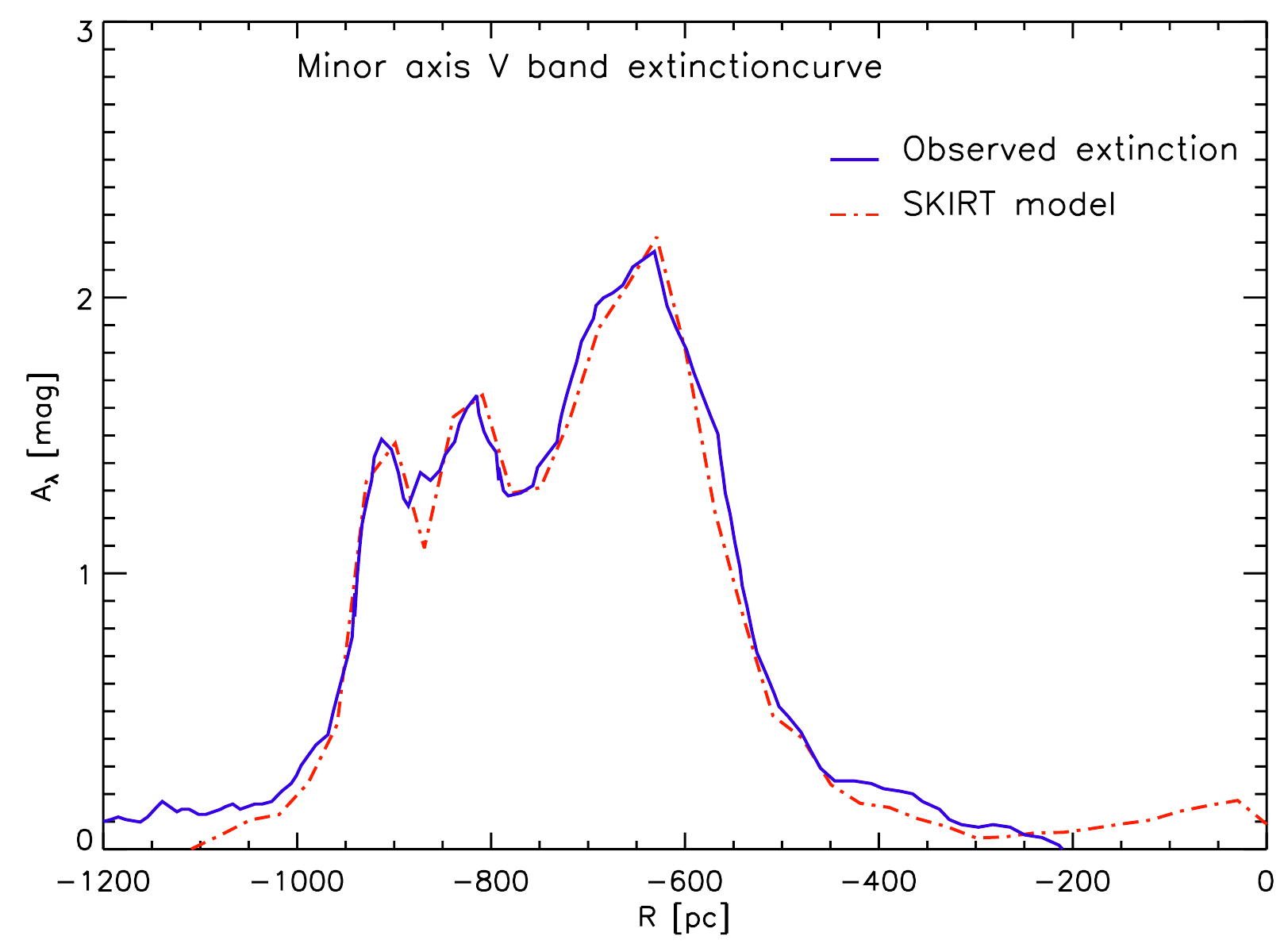} 
\includegraphics[width=0.42\textwidth]{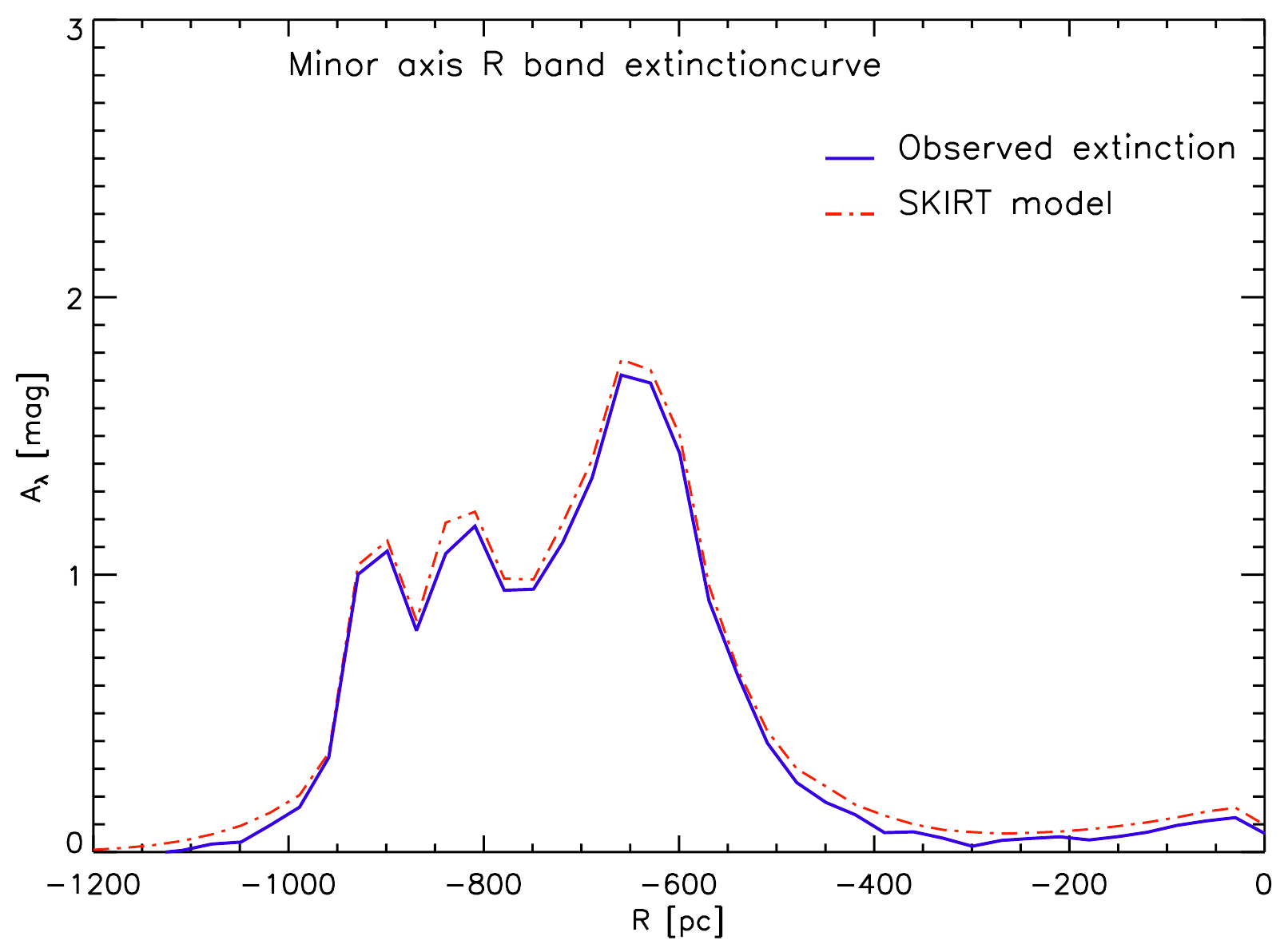} 
\caption{The modeled SED and the minor-axis extinction profiles bands for the standard model. The top panel represents the total SED as obtained with the SKIRT model (solid black line) overlaid with the observed fluxes (see Table \ref{dataptn}). The contribution of the dust to the SED is also shown. The central and bottom panels show the minor-axis extinction profile in the $V$ and $R_{\text{C}}$, where the solid blue and dashed red lines represent the observed and modeled extinction profiles, respectively.}  \label{StandardModel-1.pdf}
\end{figure}

\begin{figure*} \centering 
\includegraphics[width=0.95\textwidth]{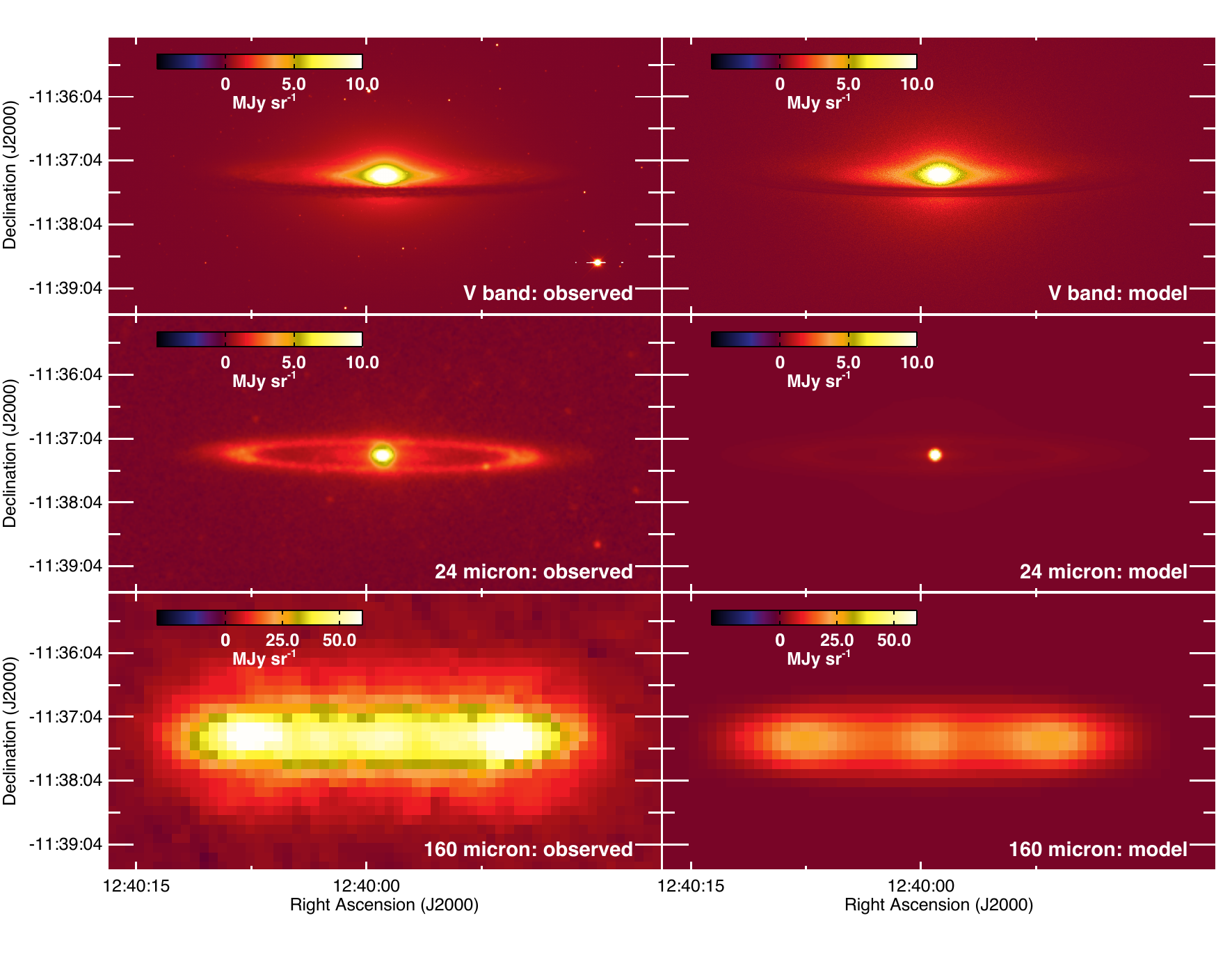}
\caption{A comparison of the observed and modeled images for the Sombrero galaxy for the standard model. Top, central and bottom rows represent images in the optical $V$ band, the MIPS 24 and MIPS 160 bands. {Images in a specific band are scaled accordingly for both observed and modeled data, allowing an immediate visual comparative inspection.}}  \label{StandardModel.pdf} 
\end{figure*}

The spatial distribution of dust in the Sombrero ring was optimized to fit the minor axis extinction profiles in both the $V$ and $R_{C}$ band.  Table \ref{paragauss} summarizes the parameter values for the radial distance $R_{i}$, the width $\sigma_{i}$ of the Gaussian function and the relative mass contribution $M_{i}$/$M_{\text{d}}$ for each Gaussian component. The relative mass contribution $M_{i}$/$M_{\text{d}}$ is in this case a dimensionless number, representative for the relative fraction of the total dust content in the Sombrero ring.  The total amount of dust in this basic model is found to be $M_{\text{d}}$ = $7.9$ $\times$ $10^{6}$ M$_{\odot}$, of which $M_{\text{d}}$ = $7.5$ $\times$ $10^{6}$ M$_{\odot}$ resides in the dust ring and the other $M_{\text{d}}$ = $4.1$ $\times$ $10^{5}$ M$_{\odot}$ is distributed in the inner disk.  Figures~{\ref{StandardModel-1.pdf}} and {\ref{StandardModel.pdf}} show the ability of this basic model to reproduce optical data of the Sombrero galaxy. In the upper panel of Figure~{\ref{StandardModel-1.pdf}}, the observed SED fluxes (see Table \ref{dataptn}) are overlaid on the SED model.  The middle and bottom rows represent the minor axis extinction profiles in the $V$ and $R_{\text{C}}$ band.  Figure~{\ref{StandardModel.pdf}} represent the images in the $V$, MIPS 24 $\mu$m and MIPS 160 $\mu$m wavebands, where the left and right panels correspond to the observed data and model images, respectively.

For this standard model that only accounts for the stellar contribution from an old stellar population, we examine the dust energy balance in M104.  When comparing the FIR/submm emission predicted from radiative transfer calculations to the observed quantities, we find a large discrepancy between the radiative transfer model and FIR/submm observations. Our standard model underestimates the true FIR/submm SED emission by a factor of $\sim$ 3. This lack of IR emission is also obvious if we look at the 24 and 160 $\mu$m SKIRT model images (see Figure \ref{StandardModel.pdf}, two bottom rows).  Although the basic model could reproduce the optical data, modifications to this model are necessary to eliminate the remaining discrepancy in the IR emission.

\begin{table}
\caption{The parameters of the 4 Gaussian components}
\begin{center}
\label{paragauss}
\begin{tabular}{@{}llll}
\hline \hline 
Standard model \\
$M_{\text{d,ring}}$ = 7.5 $\times$ 10$^{6}$ M$_{\odot}$ \\
\hline \hline
Gaussian component & R$_{i}$ & $\sigma_{i}$ & $M_{i}$/$M_{\text{d}}$\\
\hline
1 & 6100 & 480 & 0.30 \\
2 & 7900 & 250 & 0.18 \\
3 & 8750 & 180 & 0.27 \\
4 & 6850 & 1500 & 0.25 \\
\hline 
\hline \hline 
Model with embedded SF \\
$M_{\text{d,ring}}$ = 7.8 $\times$ 10$^{6}$ M$_{\odot}$ \\
\hline \hline
Gaussian component & R$_{i}$ & $\sigma_{i}$ & $M_{i}$/$M_{\text{d}}$\\
\hline
1 & 6100 & 480 & 0.33 \\
2 & 7900 & 250 & 0.18 \\
3 & 8750 & 180 & 0.26 \\
4 & 6850 & 1500 & 0.23 \\
\hline 
\end{tabular}
\end{center}
\end{table}

\subsection{Model with embedded star formation}
\label{SFModel.sec}

Relying on the large inconsistency between model and observations, we require a substantial modification of the standard model in the previous section to account for the discrepancies in dust emission. A possible explanation for the energy balance problem in the Sombrero galaxy is the presence of young stars embedded in heavily obscured dust clouds.  This additional source of dust heating will provide a significant boost to the thermal dust heating, while their locally absorbed radiation only has a minor contribution to the global extinction of starlight.  The presence of an additional component of embedded stars in the dust lane was already claimed by \citet{1991A&A...241...42K} to explain the inconsistencies between his modeled and observed $B$ and $V$ band extinction profiles.

Here, we investigate whether an additional component of young stars in our model can account for the missing dust emission. To obtain a self-consistent model for the Sombrero galaxy, we not only require fitting the thermal emission at FIR/submm wavelengths, but also the optical SED, $V$ and $R_{C}$ band images and extinction profiles still need to be reproduced.  Considering that young stars supply an additional dust heating source, the amount of dust will need reconsideration in order to still account for the observed extinction profiles. In response to an additional component of young stars, the extinction profile in the $V$ band might decrease (depending on the amount of young stars).  Although the emission from young stars will augment the number of absorption events taken place, the total amount of emitted $V$ band photons grows even more, causing a drop in the relative amount of extinction. Nevertheless we expect the changes to be small in the $V$ band and almost negligible in the $R_{C}$ band, since the bulk of emission from these young stars peaks at shorter wavelengths (UV).

We now explore the validity of models constructed from our standard model complemented with a young stellar population.  Therefore, we supplement the standard model from the previous section with a star formation component both in the dust ring and the inner disk.  \citet{Vlahakis} concluded that the dominant heating sources for the dust ring are located on the inside of the ring, based on the more extended dust ring at 870 $\mu$m and 1.2 mm compared to the emission in the wavelength range from 5.7 to 70 $\mu$m. Therefore, we insert $\sim$ 75$\%$ of this young stellar population in {the first component of the dust ring} (located closest to the nucleus). The remaining 25\% resides in the second density peak of the dust ring. This distribution more or less corresponds to the ratio of the dust masses in the first and second Gaussian component describing the dust lane in the Sombrero galaxy. For the distribution of star forming regions in M104 in the inner disk, we assume the same model describing the spatial distribution of dust in the disk.

The SED of this young stellar population is parametrized as a Starburst SED, generated from data of the Starburst99 library \citep{1999ApJS..123....3L}, with the same metallicity (Z = 0.016) as the old stellar population (10 Gyr) in the Sombrero galaxy. This starburst SSP represents a stellar population with a constant, continuous star formation, active during the past 100 Myr. 

To achieve the best fitting model that also accounts for the excess in the FIR/submm emission, we redo the radiative transfer modeling for the standard model, complemented with a SF component in the ring and the disk. Constructing a new library of radiative transfer models, we determine the parameters of the best fitting model through a least-square fitting procedure to the SED fluxes and images at 24 and 70 $\mu$m, because the bulk of thermal dust emission heated by this young stellar population will peak at these shorter IR wavelengths.

The best fitting model complements the standard model with a stellar component of modest SF activity in the inner part of the dust ring (SFR $\sim$ 0.05 M$_{\odot}$ yr$^{-1}$), while a somewhat higher star formation activity resides in the inner disk (SFR $\sim$ 0.21 M$_{\odot}$ yr$^{-1}$). The dust mass in the dust ring was a little increased ($M_{\text{d}}$ = 7.8 $\times$ 10$^{6}$ M$_{\odot}$) to maintain the resemblance between the modeled and observed $V$ band extinction curve, while the dust mass in the inner disk did not require an additional dust mass.  The relative distribution of dust in {the different components of the ring} needed some fine-tuning due to the presence of a young stellar component (see Table \ref{paragauss}).

\begin{figure} 
\centering 
\includegraphics[width=0.45\textwidth]{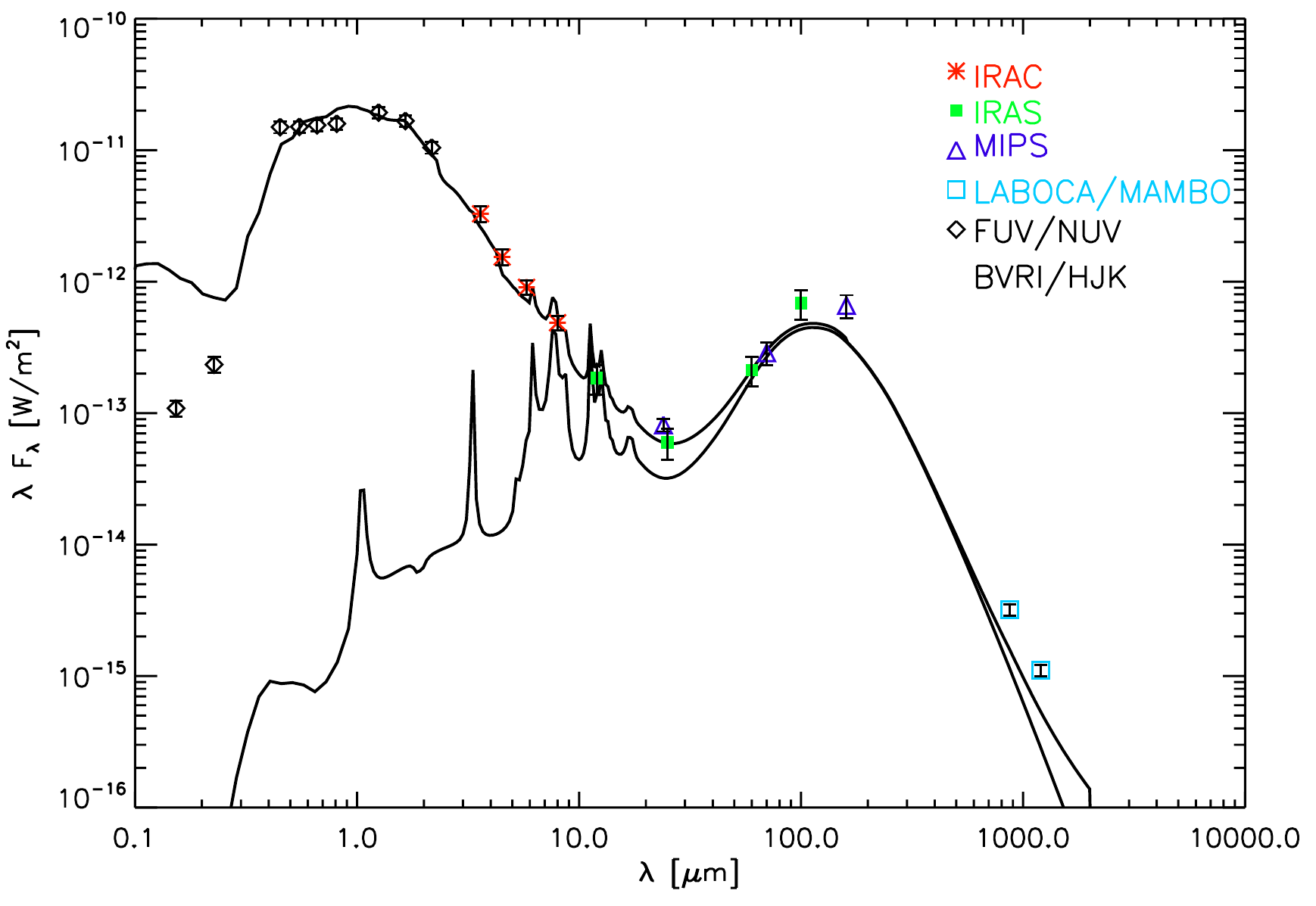} 
\includegraphics[width=0.42\textwidth]{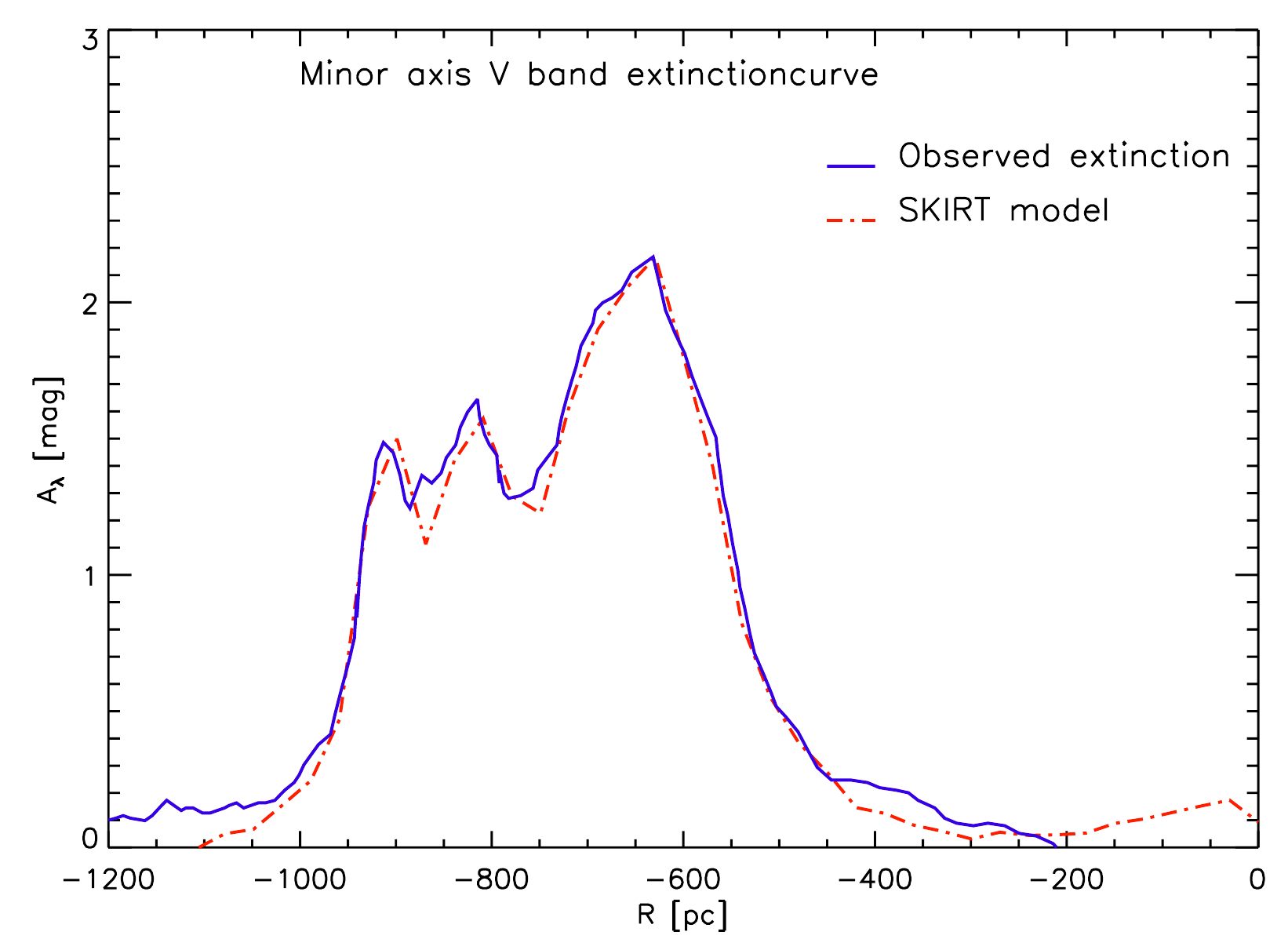} 
\includegraphics[width=0.42\textwidth]{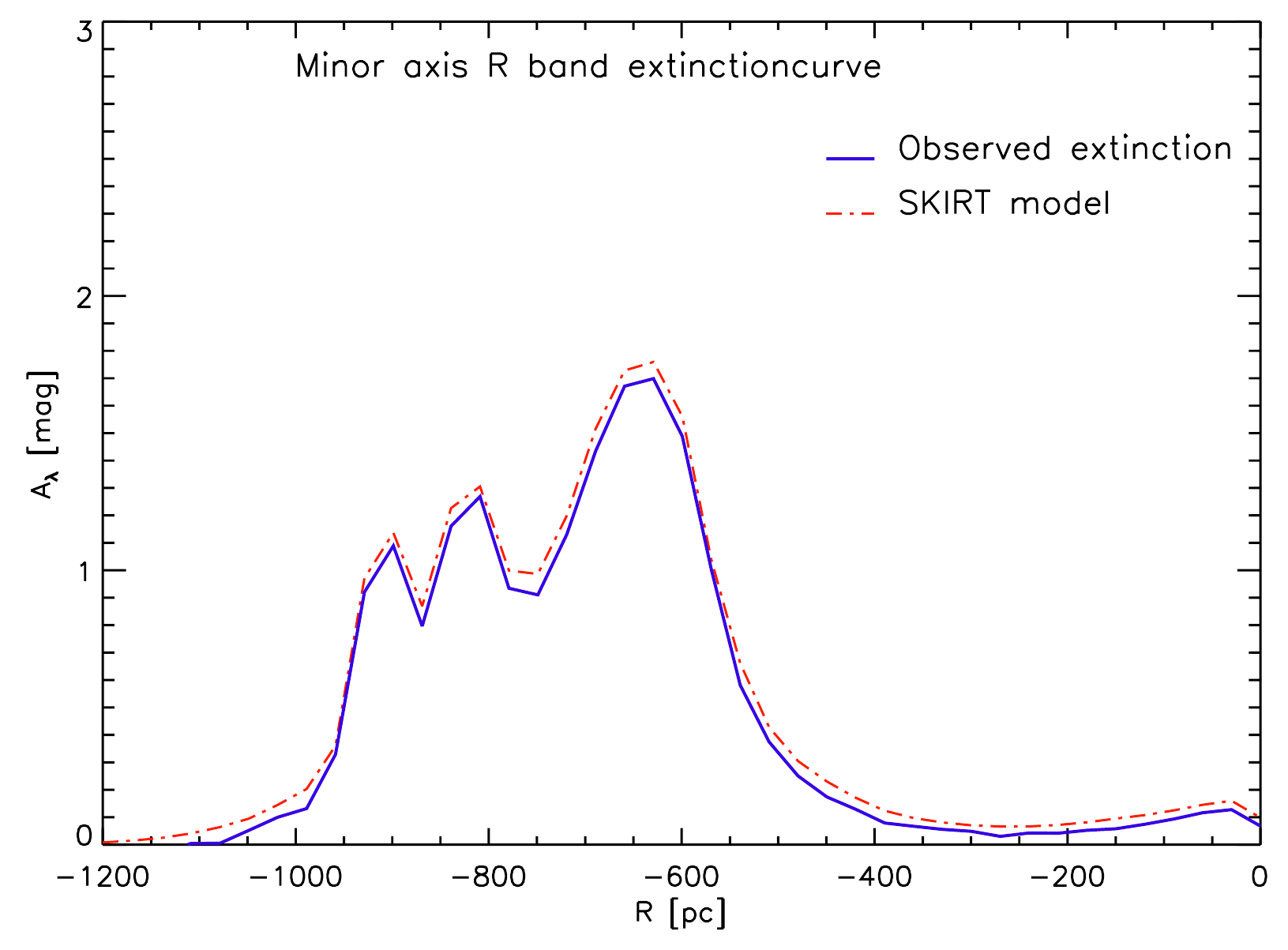} 
\caption{Same as Figure~{\ref{StandardModel-1.pdf}}, but for the model with embedded star formation.}  \label{SFModel-1.pdf}
\end{figure}

\begin{figure*} \centering 
\includegraphics[width=0.95\textwidth]{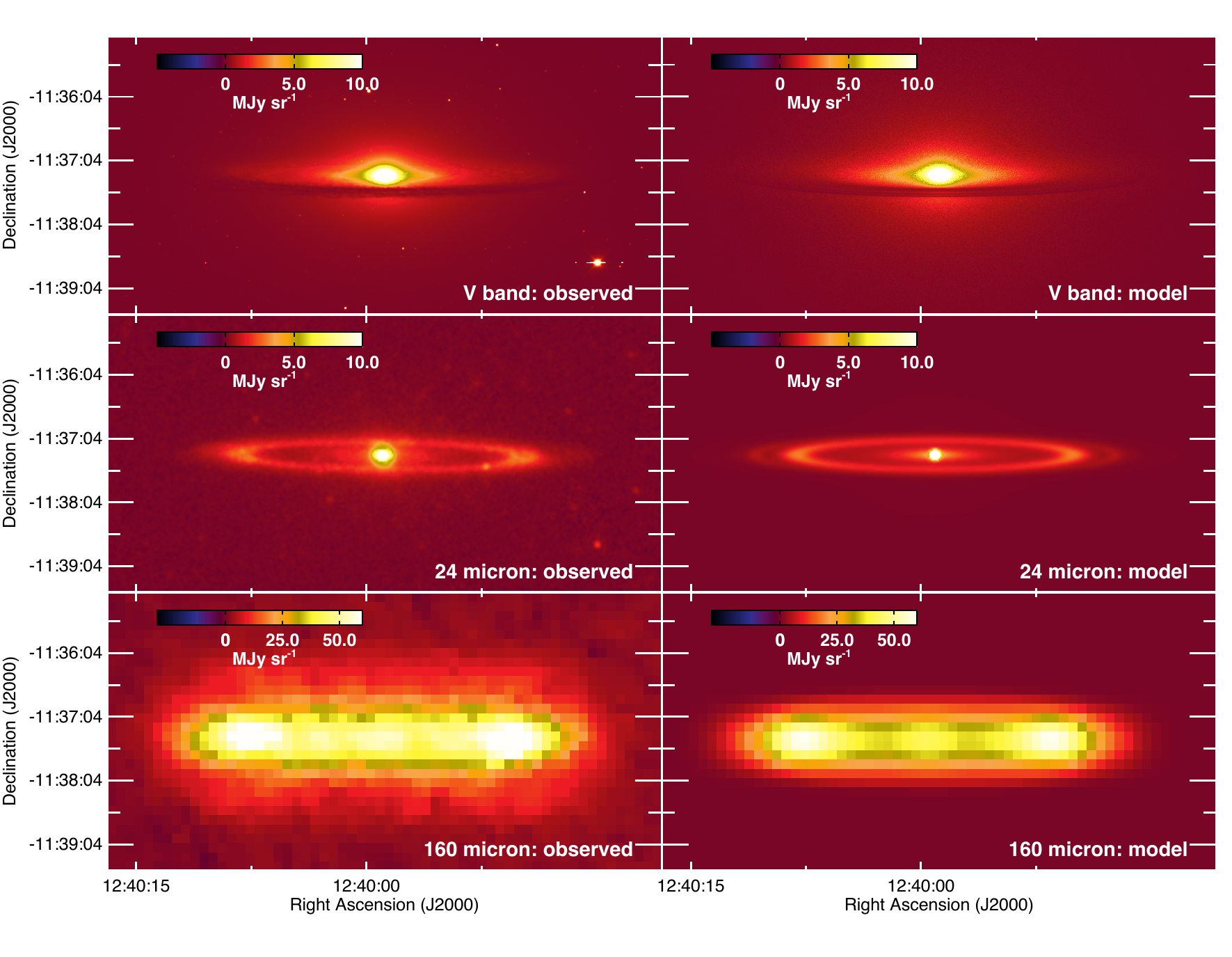}
\caption{Same as Figure~{\ref{StandardModel.pdf}}, but for the model with embedded star formation. {Also the scaling of the image graphics was adopted from Figure \ref{StandardModel.pdf}.}}  \label{SFModel.pdf} 
\end{figure*}

{Considering that the Sombrero ring contains the majority of dust mass, it is rather surprising that the bulk of recent star formation activity occurred in the inner disk, close to the nucleus. On the other hand, a recent image decomposition analysis based on IRAC data supports a more disk-like bulge structure, distinct from the prominent stellar halo in M104 \citep{2010MNRAS.403.2053G}. In view of this scenario, the presence of a young stellar population in the plane of the disk-like bulge already seems more plausible.}

This model with embedded SF is able to fit the 24 and 70 $\mu$m observations (see top panel in Figure~{\ref{SFModel-1.pdf}}). The young stars also provide a heating source for very small grains and PAH molecules. Indeed, the typical peaks characteristic for PAH emission show up in mid-infrared wavebands in the model SED. This result is consistent with the low-resolution IRS spectrum taken from the ring in M104, which shows the most prominent PAH feature in the 11.3 $\mu$m band, consistent with star formation activity in the ring \citep{Bendo}.

Although the observed FUV and NUV fluxes {do not agree with} this specific amount of stellar activity, this dissimilarity is likely a consequence of the highly obscured nature of the embedding young stars accounting for a significant part of the dust heating of grains emitting at 24 and 70 $\mu$m. 
The total star formation activity (SFR $\sim$ 0.26 M$_{\odot}$ yr$^{-1}$) in our model agrees well with the SFR estimate obtained from H$\alpha$ and radio observations of the Sombrero galaxy in \citet{2009ApJ...703.1672K} and is of the same order of magnitude as other SFR estimates based on a {suite} of other diagnostics (see Table \ref{SFR}). This makes us confident that the star formation activity in this model was not overestimated.
Table \ref{SFR} summarizes the predictions for the SFR from different diagnostics and also mentions the reference to the applied SFR relation. All values of the SFR were adjusted to the \citet{2001MNRAS.322..231K} IMF.  

While this model with embedded star formation can account for the dust emission at infrared wavelengths shortwards of 100 $\mu$m, the SED and MIPS 160 $\mu$m image at the bottom row in Figure~\ref{SFModel.pdf} clearly show that the IR emission at longer wavelengths ($>100~\mu$m) is still significantly underestimated.

\begin{table}
\caption{The different SFR tracers and estimates for M104.}
\label{SFR}
\begin{tabular}{@{}lll}
\hline \hline \\
SFR tracer(s) & SFR relation & SFR [M$_{\odot}$ yr$^{-1}$]\\
\hline
FUV & \citet{Salim} & 0.04 \\
24$\mu$m & \citet{2009ApJ...692..556R} & 0.07  \\
FUV+24$\mu$m & \citet{2008ApJ...686..155Z}  (ext. corr.) & 0.15  \\
& \citet{2009ApJ...703.1672K}  (SFR rel.) & \\
FUV+TIR & \citet{2008MNRAS.386.1157C} (ext. corr.) & 0.19 \\
& \citet{2009ApJ...703.1672K}  (SFR rel.) & \\
H$\alpha$+8$\mu$m & \citet{2009ApJ...703.1672K} & 0.14 \\
H$\alpha$+24$\mu$m & \citet{2009ApJ...703.1672K} & 0.07\\
H$\alpha$+TIR & \citet{2009ApJ...703.1672K} & 0.16 \\
H$\alpha$+1.4GHz & \citet{2009ApJ...703.1672K} & 0.26 (total)\\
& & 0.06 (disk) \\
\hline 
\end{tabular}
\end{table}

\section{Discussion}

The most important result from our analysis is the inability to reproduce the IR/submm emission at wavelengths longwards of 100 $\mu$m. The emission from dust grains in our model with embedded star formation underestimates the observed IR/submm emission by a factor of $\sim$ 3. A similar energy balance problem has been encountered in other spiral galaxies. In this section we discuss the most likely explanation for the discrepancy in the energy balance of the Sombrero galaxy and compare these results with energy balance studies in other objects.

Several edge-on spirals have been the target of a detailed energy balance analysis, among which NGC\,891 stands out as the most extensively studied object.  In general, those studies indicate that 30$ \%$ of the UV/optical photons is absorbed and re-emitted in the infrared, while the optical accounts for the absorption of only 10$\%$ of the stellar radiation \citep{Popescu, 2001A&A...372..775M, 2002MNRAS.335L..41P, Alton, Dasyra, 2010A&A...518L..39B}. This inconsistency is referred to as the `dust energy balance problem'.

\citet{Popescu, 2011A&A...527A.109P} and \citet{2001A&A...372..775M} tried to explain the discrepancy in the energy budget by including a second thin diffuse dust disk, associated with a young stellar population. Although this additional dust accounts for the missing observed FIR/submm emission, \citet{Dasyra} and \citet{Bianchi2007} show that this second disk scenario contradicts with \textit{K} band observations. Alternatively, a secondary dust component can be added in optically thick clumps distributed throughout the dust disk. Neglecting a clumpy dust structure can result in an underestimate of the optical dust mass by 40-50$\%$ \citep{1999AAS...194.7121W,2000A&A...356..795A,2002A&A...384..866M}. These clumps can either have embedded star forming regions or have no associated sources. The latter clumps are referred to as `quiescent'. \citet{2008A&A...490..461B} shows that a clumpy structure, encompassing half of the dust mass (compared to the diffuse dust component) and embedded sources, is able to solve the energy balance problem in NGC\,891. Also the results in \citet{2010A&A...518L..39B} support the presence of heavily obscured star formation in clumpy dust clouds, to account for the missing infrared radiation in UGC\,4754.

An alternative explanation that does not require the assumption of an additional dust component was explored in \citet{2000A&A...356..795A, Alton} and \citet{Dasyra}. They argue that an increased emissivity in the submm wavelength range compared to Galactic values, both in a diffuse component and in denser environments, also offers a possible solution to the energy balance problem. Assuming an increased emissivity by a factor of 4 and 1.5 at 850 $\mu$m and 1.2mm, respectively, compared to the model prediction from \citet{DraineLee} and high latitude observations in the Milky Way, allows \citet{Alton} to account for the discrepancy in the energy balance in three edge-on spirals (NGC\, 4013, NGC\,4565, NGC\,5907).  Although many authors have explored different explanations for this excess in dust emission, due to a lack of high-resolution multi-wavelength data one has not come to a general consensus regarding the cause of this contradiction in the dust energy balance for several edge-on spirals.

In the same manner that previous energy balance studies have shown to underestimate the IR emission by a factor of $\sim$ 3, the standard model in this work, which was constructed from a fitting procedure to the available optical data, lacks a similar amount of dust emission in the infrared wavebands.

Relying on the agreement of our model with the SFR calibrations and 24 and 70 $\mu$m observations, we argue that an additional component of heavily extinguished star formation will not bring a solution to the remaining discrepancy.  Rather than embedded localized sources, it seems that either an additional dust component, not associated to star forming regions, or dust grains with a higher emissivity at FIR/submm wavelengths, compared to Galactic values, need to be invoked to explain the discrepancy in the energy budget at longer wavelengths ($>100~\mu$m).

The former assumption of a dust component distributed in clumps throughout the dust ring is a reasonable assumption. \citet{Emsellem} already inferred a clumpy dust structure from his absorption $A_{V}^{\text{eff}}$ map in the southern part of M104 (see Figure 4 in his paper). 
Also, high resolution optical HST data \citep{2003AAS...20311611C} of the Sombrero galaxy hint at a clumpy dust structure.

In late-type galaxies the discrepancy between model and IR observations mainly arises at shorter IR wavelengths (NGC\,891: \citealt{2008A&A...490..461B} ; UGC 4754: \citealt{2010A&A...518L..39B}), where the warm dust component dominates the IR emission spectrum, and, hence, the presence of embedded star-forming clouds seems to be the most likely explanation for the dust energy balance problem in those objects.
In case of the Sombrero galaxy, an early-type spiral (Sa), the majority of compact clumps are not associated to localized sources and, therefore, have not been triggered to for stars (yet).
These compact clumps are optically thick to the diffuse UV and optical radiation field and, therefore, will not have a substantial influence on the observed extinction profile, because most of the diffuse UV/optical radiation from young stellar objects is locally absorbed. Since these quiescent clumps are not associated to any source in a star forming region, we could consider them as dust clouds for which the conditions are not appropriate to initiate star formation. Because these clouds might still start forming stars, we expect the density of dust in these clumps to be very high. Therefore, these compact structures could contain a substantial amount of dust in addition to the diffuse dust component, which is distributed more smoothly among the dust lane. 

\begin{figure}
\centering
\includegraphics[width=85mm]{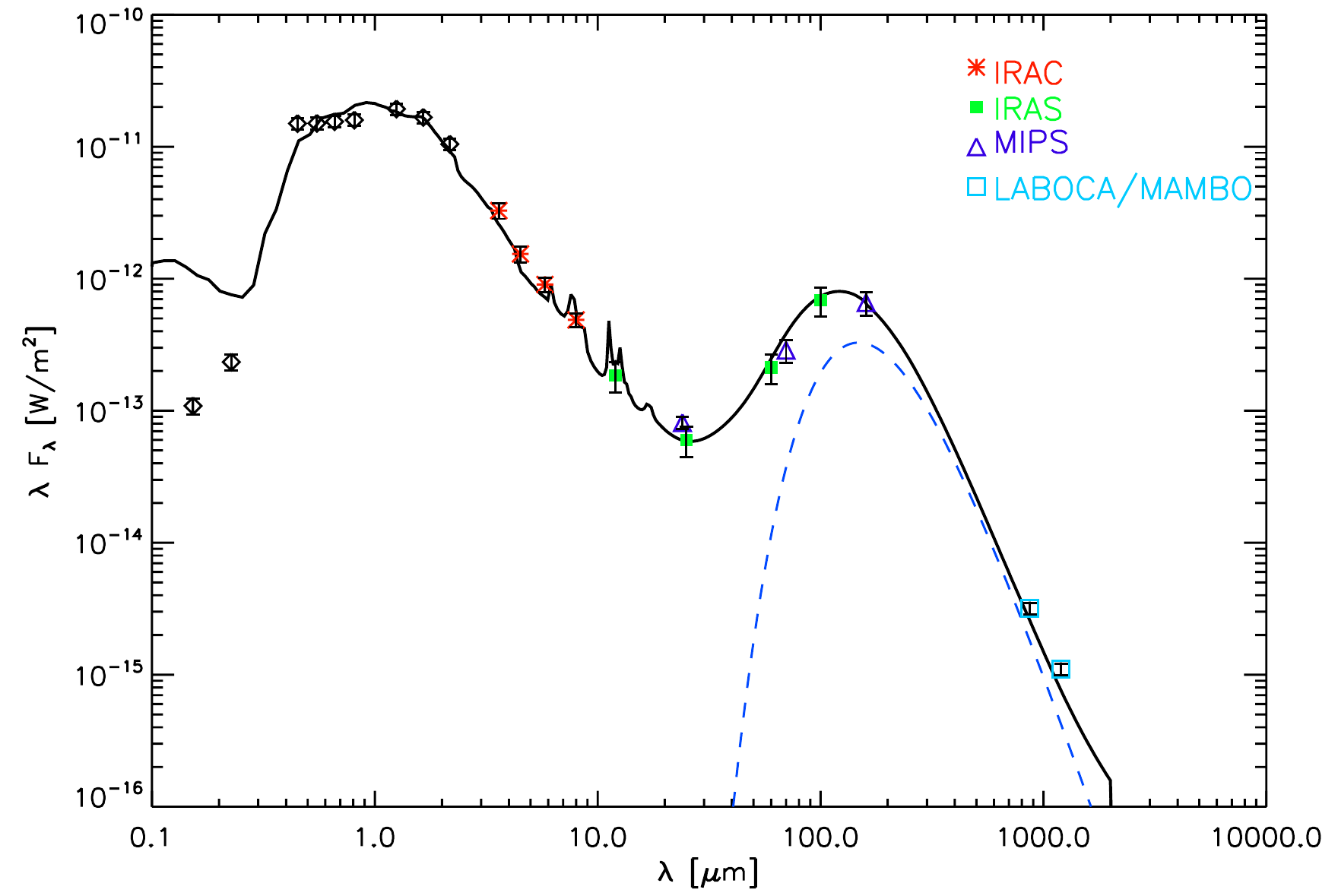}
\caption{Spectral energy distribution (SED) for the Sombrero with an embedded star formation component, supplemented with a dust component $M_{\text{d}}$~=~2.0 $\times$ $10^{7}$ M$_{\odot}$, distributed in clumps. The solid line represents the modeled SED. The modified black-body, modeling the additional dust component in compact clumps, is shown by the blue dashed-dotted line. The symbols correspond to observations in the optical and infrared spectrum.}
\label{ExtraSED.pdf}
\end{figure}

Given the low star formation activity in the Sombrero galaxy (SFR $\sim$ 0.26 M$_{\odot}$ yr$^{-1}$) compared to late-type galaxies such as NGC\,891 (SFR $\sim$ 3 M$_{\odot}$ yr$^{-1}$, \citealt{Popescu, 2008A&A...490..461B}) and NGC\,4565 (SFR $\sim$ 0.7 M$_{\odot}$ yr$^{-1}$, \citealt{2006PASP..118.1098W}), it is not surprising that M104 overall would contain more quiescent dust clumps, with no associated embedded sources. To achieve an estimate of the total amount of dust residing in dense clumps, we calculate the dust mass required to explain the dust energy balance problem in M104. Therefore, we subtract the emission from the model with embedded star formation from the observed fluxes at 100, 160, 870 and 1200 $\mu$m, respectively. By fitting a single-component modified blackbody to the residual fluxes, we obtain a dust mass of $M_{\text{d}}$ = 2.0 $\times$ 10$^{7}$ M$_{\odot}$ at a temperature of $T_{d}$ = 16.2 K.  Making the assumption of an additional dust component distributed in quiescent compact clumps throughout the dust lane of the Sombrero galaxy can account for the IR emission, also at wavelengths longer than 100 $\mu$m (see Figure \ref{ExtraSED.pdf}).

\citet{2008A&A...490..461B} already showed that only a dust distribution for which half of the dust is distributed in clumps can explain the energy balance in NGC\,891. In the Sombrero galaxy, the total amount of dust residing in clumps is estimated to exceed the amount of diffusely distributed dust and account for about three-quarters of the total dust content ($M_{\text{d}}$ = 2.8 $\times$ 10$^{7}$ M$_{\odot}$). The dust mass in our model is a bit higher than previous estimates of the dust content in M104, when scaled to the same distance (9.2 Mpc). \citet{Emsellem} and \citet{Bendo} inferred a dust mass ($M_{\text{d}}$ = 7.9 $\times$ 10$^{6}$ M$_{\odot}$) comparable to the dust content in our standard model. When taking into account observations at longer wavelengths, a total dust content of 1.3 $\times$ 10$^{7}$ M$_{\odot}$ and 1.5 $\times$ 10$^{7}$ M$_{\odot}$ with dust temperatures ($T_{d}$ = 22 K and 18.4 K, respectively) was found in \citet{2006A&A...448..133K} and \citet{Vlahakis} from radio and submm/mm observations, respectively. Recent Herschel observations have estimated a dust mass of $M_{\text{d}}$ $\sim$ 8.1 $\times$ 10$^{6}$ M$_{\odot}$ at a temperature of $T_{d}$ = 22 K \citep{Skibba2011}. 
The small dissimilarity with the higher dust content in our self-consistent Sombrero dust model is largely due to the lower dust temperature for the majority of the dust grains in our model ($T_{d}$ = 16.2 K).

\section{Conclusions}

In this work, we have studied the energy balance of the Sombrero galaxy in a full radiative transfer analysis, including scattering, absorption and thermal dust re-emission.  The Sombrero galaxy benefits from its 84$^{\circ}$ inclination in such a way that the inner disk of the Sombrero galaxy can be revealed, while most details, and hereby information about the dust and stellar properties in the disk, vanish when integrating along the line-of-sight in more highly inclined objects. Complementing this favorable inclination with the very symmetric dust lane in M104 and the availability of high-resolution FIR data, the Sombrero galaxy is considered an ideal study object for a detailed study of the energy balance.

From radiative transfer simulations, we obtain a self-consistent model for the Sombrero galaxy, which is able to account for the observed stellar SED, the extinction of dust particles in the dust lane as well as the images at optical and MIR wavelengths. This best fitting model assumes a low-activity star forming stellar component both in the ring (SFR $\sim$ 0.05 M$_{\odot}$ yr$^{-1}$) and inner disk (SFR $\sim$ 0.21 M$_{\odot}$ yr$^{-1}$) of the Sombrero galaxy to account for the excess in emission at 24 and 70 $\mu$m.  Although this model with embedded star formation is in agreement with observations of the dust component heated by young stars, it still underestimates the dust emission in wavebands longwards of 100 $\mu$m.  When assuming the presence of an additional dust component, accounting for three-quarters of the total dust content ($M_{\text{d}}$ $\sim$ 2.8 $\times$ 10$^{7}$ M$_{\odot}$), distributed in quiescent compact clumps throughout the dust ring, we can account for the observed excess at FIR/submm wavelengths.  The presence of a clumpy dust structure is supported by high-resolution HST data, revealing the clumpiness of dust in the Sombrero dust ring.  In correspondence to the lower star formation activity in early-type galaxies compared to late-type objects, the dust clumps in the Sombrero galaxy do not seem to be associated to recent star formation.

Although a clumpy dust structure is a reasonable assumption, we are not capable of completely ruling out the possibility that part of the discrepancy might be caused by a dust component composed of grains with a higher emissivity in the submm wave bands, compared to Galactic properties.
{Future studies of the energy balance in thirty galaxies, for which ongoing \textit{Herschel} observations (HEROES, NHEMESES and FRIEDL programs) soon become available, will significantly improve the sample statistics and, hopefully, shed light on the primary cause (additional dust component or dust with different emissivity properties) for the crisis in the energy budget in those objects.}

\section*{Acknowledgements}

{We thank the referee for his/her comments on the paper, which helped to improve the paper. We thank Dimitri Gadotti and Karl Gordon for stimulating discussions, and George Bendo for sharing his expertise in image plotting. IDL, MB and JF kindly acknowledge financial support from the Belgian Science Policy Office (BELSPO), MB and JV acknowledge the support of the Flemish Fund for Scientific Research (FWO-Vlaanderen).}\\

\appendix  

\bsp  

\label{lastpage}


\begin{thebibliography}{}
\bibitem[\protect\citeauthoryear{Ajhar et al.}{1997}]{1997AJ....114..626A} Ajhar, E.~A., Lauer, T.~R., Tonry, J.~L., Blakeslee, J.~P., Dressler, A., Holtzman, J.~A., \& Postman, M. 1997, AJ, 114, 626 
\bibitem[\protect\citeauthoryear{Alton et al.}{2000}]{2000A&A...356..795A} Alton, P.~B., Xilouris, E.~M., Bianchi, S., Davies, J., \& Kylafis, N. 2000, A\&A, 356, 795
\bibitem[\protect\citeauthoryear{Alton~et al.}{2004}]{Alton} Alton~P.~B. et al., 2004, A\&A, 425, 109
\bibitem[\protect\citeauthoryear{Baes \& Dejonghe}{2002}]{2002MNRAS.335..441B} Baes, M., \& Dejonghe, H. 2002, MNRAS, 335, 441 
\bibitem[\protect\citeauthoryear{Baes et al.}{2003}]{2003MNRAS.343.1081B} Baes, M., et al. 2003, MNRAS, 343, 1081
\bibitem[\protect\citeauthoryear{Baes et al.}{2005a}]{2005AIPC..761...27B} Baes, M., Dejonghe, H., \& Davies, J.~I.\ 2005, The Spectral Energy Distributions of Gas-Rich Galaxies: Confronting Models with Data, 761, 27 
\bibitem[\protect\citeauthoryear{Baes et al.}{2005b}]{2005NewA...10..523B} Baes, M., Stamatellos, D., Davies, J.~I., Whitworth, A.~P., Sabatini, S., Roberts, S., Linder, S.~M., \& Evans, R. 2005, NA, 10, 523
\bibitem[\protect\citeauthoryear{Baes et al.}{2010}]{2010A&A...518L..39B} Baes, M., et al. 2010, A\&A, 518, L39 
\bibitem[\protect\citeauthoryear{Baes et al.}{2011}]{Baes2011} Baes M., Verstappen J., De Looze I., Fritz J., Saftly W., Vidal PŽrez E., Stalevski M., Valcke S., 2011, ApJS, in press (arXiv:1108.5056)
\bibitem[\protect\citeauthoryear{Bajaja et al.}{1984}]{1984A&A...141..309B} Bajaja, E., van der Burg, G., Faber, S.~M., Gallagher, J.~S., Knapp, G.~R., \& Shane, W.~W. 1984, A\&A, 141, 309
\bibitem[\protect\citeauthoryear{Bajaja~et al.}{1988}]{Bajaja1} Bajaja E., 1988, A\&A, 202, 35
\bibitem[\protect\citeauthoryear{Bajaja et al.}{1991}]{1991A&A...241..411B} Bajaja, E., Krause, M., Wielebinski, R., \& Dettmar, R.-J. 1991, A\&A, 241, 411 
\bibitem[\protect\citeauthoryear{Bendo~et al.}{2006}]{Bendo} Bendo G., 2006, AJ, 645, 134 
\bibitem[\protect\citeauthoryear{Bianchi}{1999}]{Bianchi1999} Bianchi S., Davies J.I., Alton P.B., 1999, A\&A, 344, L1 
\bibitem[\protect\citeauthoryear{Bianchi}{2007}]{Bianchi2007} Bianchi S., 2007, A\&A, 471, 765 
\bibitem[\protect\citeauthoryear{Bianchi}{2008}]{2008A&A...490..461B} Bianchi, S. 2008, A\&A, 490, 461 
\bibitem[\protect\citeauthoryear{Bot et al.}{2010}]{2010A&A...524A..52B} Bot, C., et al. 2010, A\&A, 524, A52 
\bibitem[\protect\citeauthoryear{Burkert \& Tremaine}{2010}]{2010ApJ...720..516B} Burkert, A., \& Tremaine, S. 2010, ApJ, 720, 516 
\bibitem[\protect\citeauthoryear{Burkhead}{1986}]{1986AJ.....91..777B} Burkhead, M.~S.1986, AJ, 91, 777 
\bibitem[\protect\citeauthoryear{Calzetti~et al.}{2007}]{Calzetti} Calzetti~D. et al., 2007, AJ, 666, 870
\bibitem[\protect\citeauthoryear{Caplan \& Deharveng}{1986}]{1986A&A...155..297C} Caplan, J., \& Deharveng, L. 1986, A\&A, 155, 297 
\bibitem[\protect\citeauthoryear{Cappellari}{2002}]{2002MNRAS.333..400C} Cappellari, M. 2002, MNRAS, 333, 400 
\bibitem[\protect\citeauthoryear{Cockcroft et al.}{2009}]{2009AJ....138..758C} Cockcroft, R., Harris, W.~E., Wehner, E.~M.~H., Whitmore, B.~C., \& Rothberg, B. 2009, AJ, 138, 758 
\bibitem[\protect\citeauthoryear{Compi{\`e}gne et al.}{2011}]{2011A&A...525A.103C} Compi{\`e}gne, M., et al. 2011, A\&A, 525, A103 
\bibitem[\protect\citeauthoryear{Cortese et al.}{2008}]{2008MNRAS.386.1157C} Cortese, L., Boselli, A., Franzetti, P., Decarli, R., Gavazzi, G., Boissier, S., \& Buat, V. 2008, MNRAS, 386, 1157
\bibitem[\protect\citeauthoryear{Christian et al.}{2003}]{2003AAS...20311611C} Christian, C.~A., Bond, H.~E., Frattare, L.~M., Hamilton, F., Levay, Z.~G., Knoll, K.~S., \& Royle, T. 2003, Bulletin of the American Astronomical Society, 35, $\#$116.11 
\bibitem[\protect\citeauthoryear{Dale~et al.}{2007}]{Dale} Dale~D.~A. et al., 2007, AJ, 655, 863 
\bibitem[\protect\citeauthoryear{Dale et al.}{2009}]{2009ApJ...703..517D} Dale, D.~A., et al. 2009, ApJ, 703, 517
\bibitem[\protect\citeauthoryear{Dasyra~et al.}{2005}]{Dasyra} Dasyra~K.M., Xilouris~E.M., Misiriotis~A., Kylafis~N.D. et al., 2005, AIPC, 761, 197
\bibitem[\protect\citeauthoryear{De Looze et al.}{2010}]{2010A&A...518L..54D} De Looze, I., et al. 2010, A\&A, 518, L54 
\bibitem[\protect\citeauthoryear{Draine \& Lee}{1984}]{DraineLee} Draine~B.T. \& Lee~H.M., 1984, ApJ, 285, 89 
\bibitem[\protect\citeauthoryear{Draine}{2003}]{2003ARA&A..41..241D} Draine, B.~T. 2003, ARA\&A, 41, 241
\bibitem[\protect\citeauthoryear{Draine \& Li}{2007}]{2007ApJ...657..810D} Draine, B.~T., \& Li, A. 2007, ApJ, 657, 810 
\bibitem[\protect\citeauthoryear{Emsellem}{1995}]{Emsellem} Emsellem~E., 1995, A\&A, 303, 673
\bibitem[\protect\citeauthoryear{Emsellem \& Ferruit}{2000}]{2000A&A...357..111E} Emsellem, E., \& Ferruit, P. 2000, A\&A, 357, 111 
\bibitem[\protect\citeauthoryear{Finkelman}{2008}]{Finkelman} Finkelman~I. et al., 2008, MNRAS, 390, 969
\bibitem[\protect\citeauthoryear{Ford et al.}{1996}]{1996ApJ...458..455F} Ford, H.~C., Hui, X., Ciardullo, R., Jacoby, G.~H., \& Freeman, K.~C. 1996, ApJ, 458, 455 
\bibitem[\protect\citeauthoryear{Gadotti et al.}{2010}]{2010MNRAS.403.2053G} Gadotti, D.~A., Baes, M., \& Falony, S. 2010, MNRAS, 403, 2053
\bibitem[\protect\citeauthoryear{Gomez et al.}{2010}]{2010A&A...518L..45G} Gomez, H.~L., et al.\ 2010, A\&A, 518, L45 
\bibitem[\protect\citeauthoryear{Harris et al.}{2010}]{2010MNRAS.401.1965H} Harris, W.~E., Spitler, L.~R., Forbes, D.~A., \& Bailin, J. 2010, MNRAS, 401, 1965
\bibitem[\protect\citeauthoryear{Kennicutt}{1983}]{Kennicutt} Kennicutt~R.C., 1983, AJ, 272, 54
\bibitem[\protect\citeauthoryear{Kennicutt et al.}{2003}]{2003PASP..115..928K} Kennicutt, R.~C., Jr., et al. 2003, PASP, 115, 928 
\bibitem[\protect\citeauthoryear{Kennicutt et al.}{2008}]{2008ApJS..178..247K} Kennicutt, R.~C., Jr., Lee, J.~C., Funes, S.~J., Jos{\'e} G., Sakai, S., \& Akiyama, S. 2008, ApJS, 178, 247 
\bibitem[\protect\citeauthoryear{Kennicutt et al.}{2009}]{2009ApJ...703.1672K} Kennicutt, R.~C., et al. 2009, ApJ, 703, 1672  
\bibitem[\protect\citeauthoryear{Knapen et al.}{1991}]{1991A&A...241...42K} Knapen, J.~H., Hes, R., Beckman, J.~E., \& Peletier, R.~F.\ 1991, A\&A, 241, 42 
\bibitem[\protect\citeauthoryear{Kormendy et al.}{1996}]{1996ApJ...473L..91K} Kormendy, J., et al. 1996, ApJL, 473, L91 
\bibitem[\protect\citeauthoryear{Krause et al.}{2006}]{2006A&A...448..133K} Krause, M., Wielebinski, R., \& Dumke, M. 2006, A\&A, 448, 133 
\bibitem[\protect\citeauthoryear{Kroupa}{2001}]{2001MNRAS.322..231K} Kroupa, P. 2001, MNRAS, 322, 231 
\bibitem[\protect\citeauthoryear{Kuchinski et al.}{1998}]{Kuchinski} Kuchinski~L.E.., Terndrup~D.M., Gordon~K.D., Witt~A.N., 1983, AJ, 272, 54
\bibitem[\protect\citeauthoryear{Kylafis \& Bahcall}{1987}]{1987ApJ...317..637K} Kylafis, N.~D., \& Bahcall, J.~N. 1987, ApJ, 317, 637 
\bibitem[\protect\citeauthoryear{Lasker}{1979}]{Lasker} Lasker~B.M., 1979, Publ. Astron. Soc. Pacific 91, 158
\bibitem[\protect\citeauthoryear{Leitherer et al.}{1999}]{1999ApJS..123....3L} Leitherer, C., et al. 1999, ApJS, 123, 3 
\bibitem[\protect\citeauthoryear{Li et al.}{2007}]{2007MNRAS.376..960L} Li, Z., Wang, Q.~D., \& Hameed, S. 2007, MNRAS, 376, 960  
\bibitem[\protect\citeauthoryear{Li et al.}{2010}]{2010ApJ...721.1368L} Li, Z., et al. 2010, ApJ, 721, 1368 
\bibitem[\protect\citeauthoryear{Maraston}{1998}]{1998MNRAS.300..872M} Maraston, C. 1998, MNRAS, 300, 872
\bibitem[\protect\citeauthoryear{Maraston}{2005}]{2005MNRAS.362..799M} Maraston, C. 2005, MNRAS, 362, 799 
\bibitem[\protect\citeauthoryear{Matsumura \& Seki}{1989}]{1989A&A...209....8M} Matsumura, M., \& Seki, M. 1989, A\&A, 209, 8 
\bibitem[\protect\citeauthoryear{Misiriotis \& Bianchi}{2002}]{2002A&A...384..866M} Misiriotis, A., \& Bianchi, S. 2002, A\&A, 384, 866 
\bibitem[\protect\citeauthoryear{Misiriotis et al.}{2001}]{2001A&A...372..775M} Misiriotis, A., Popescu, C.~C., Tuffs, R., \& Kylafis, N.~D. 2001, A\&A, 372, 775 
\bibitem[\protect\citeauthoryear{Mould \& Spitler}{2010}]{2010ApJ...722..721M} Mould, J., \& Spitler, L. 2010, ApJ, 722, 721 
\bibitem[\protect\citeauthoryear{Mu{\~n}oz-Mateos et al.}{2009}]{2009ApJ...701.1965M} Mu{\~n}oz-Mateos, J.~C., et al. 2009, ApJ, 701, 1965 
\bibitem[\protect\citeauthoryear{Ossenkopf~V.}{1993}]{Ossenkopf1993} Ossenkopf~V., 1993, A\&A, 280, 617 
\bibitem[\protect\citeauthoryear{Ossenkopf~V., Henning~Th.}{1994}]{Ossenkopf1994} Ossenkopf~V., Henning~Th., 1994, A\&A, 291, 943 
\bibitem[\protect\citeauthoryear{Pilbratt et al.}{2010}]{2010A&A...518L...1P} Pilbratt, G.~L., et al., 2010, A\&A, 518, L1
\bibitem[\protect\citeauthoryear{Popescu~et al.}{2000}]{Popescu} Popescu~C.C., Misiriotis~A., Kylafis~N.D., Tuffs~R.J., Fischera~J., 2000, A\&A, 362, 138
\bibitem[\protect\citeauthoryear{Popescu \& Tuffs}{2002}]{2002MNRAS.335L..41P} Popescu, C.~C., \& Tuffs, R.~J. 2002, MNRAS, 335, L41 
\bibitem[\protect\citeauthoryear{Popescu et al.}{2011}]{2011A&A...527A.109P} Popescu, C.~C., Tuffs, R.~J., Dopita, M.~A., Fischera, J., Kylafis, N.~D., \& Madore, B.~F. 2011, A\&A, 527, A109 
\bibitem[\protect\citeauthoryear{Rice et al.}{1988}]{1988ApJS...68...91R} Rice, W., Lonsdale, C.~J., Soifer, B.~T., Neugebauer, G., Kopan, E.~L., Lloyd, L.~A., de Jong, T., \& Habing, H.~J. 1988, ApJS, 68, 91 
\bibitem[\protect\citeauthoryear{Rieke et al.}{2009}]{2009ApJ...692..556R} Rieke, G.~H., Alonso-Herrero, A., Weiner, B.~J., P{\'e}rez-Gonz{\'a}lez, P.~G., Blaylock, M., Donley, J.~L., \& Marcillac, D. 2009, ApJ, 692, 556
\bibitem[\protect\citeauthoryear{Salim et al.}{2007}]{Salim} Salim S. et al., 2007, ApJS, 173, 267
\bibitem[\protect\citeauthoryear{Schmitt et al.}{1997}]{1997AJ....114..592S} Schmitt, H.~R., Kinney, A.~L., Calzetti, D., \& Storchi Bergmann, T. 1997, AJ, 114, 592 
\bibitem[\protect\citeauthoryear{Schweizer}{1978}]{Schweizer} Schweizer~F., 1978, AJ, 220, 98
\bibitem[\protect\citeauthoryear{Skibba et al.}{2011}]{Skibba2011} Skibba, R.~A. et al., 2011, ApJ, in press, arXiv:1106.4022
\bibitem[\protect\citeauthoryear{Spitler et al.}{2008}]{2008MNRAS.385..361S} Spitler, L.~R., Forbes, D.~A., Strader, J., Brodie, J.~P., \& Gallagher, J.~S. 2008, MNRAS, 385, 361 
\bibitem[\protect\citeauthoryear{Stalevski et al.}{2011}]{Stalevski2011} Stalevski, M., et al. 2011, MNRAS, submitted 
\bibitem[\protect\citeauthoryear{Stepnik~et al.}{2003}]{Stepnik} Stepnik~B., Abergel~A., Bernard J.-P. et al., 2003, A\&A, 398, 551
\bibitem[\protect\citeauthoryear{Vazdekis~et al.}{1997}]{Vazdekis} Vazdekis~A., Peletier~R.~F., Beckman~J.~E.,Casuso~E., 1997, AJ, 111, 203
\bibitem[\protect\citeauthoryear{Vidal \& Baes}{2007}]{2007BaltA..16..101V} Vidal, E., \& Baes, M. 2007, Baltic Astronomy, 16, 101
\bibitem[\protect\citeauthoryear{Vidal et al.}{2011}]{Vidal2011} Vidal, E., et al. 2011, A\&A, in prep. 
\bibitem[\protect\citeauthoryear{Vlahakis~et al.}{2008}]{Vlahakis} Vlahakis~C., Baes~M., Bendo~G., Lundgren~A., 2008, A\&A, 485, L25
\bibitem[\protect\citeauthoryear{Witt~A., Thronson~H.A., Capuano~J.M.}{1992}]{Witt} Witt~A., Thronson~H.A., Capuano~J.M., 1992, AJ, 393, 611
\bibitem[\protect\citeauthoryear{Witt \& Gordon}{1999}]{1999AAS...194.7121W} Witt, A.~N., \& Gordon, K.~D. 1999, Bulletin of the American Astronomical Society, 31, 946 
\bibitem[\protect\citeauthoryear{Xilouris~et al.}{1997}]{Xilouris1997} Xilouris~E.M., Kylafis N.D., Papamastorakis J., Paleologou E.V., Haerendel G., A\&A, 325, 135
\bibitem[\protect\citeauthoryear{Xilouris~et al.}{1998}]{Xilouris1998} Xilouris~E.M., Alton P.B., Davies J.I., Kylafis N.D., Papamastorakis J., Trewhella M., A\&A, 331, 894
\bibitem[\protect\citeauthoryear{Xilouris~et al.}{1999}]{Xilouris1999} Xilouris~E.M., Byun Y.I., Kylafis N.D., Paleologou E.V., Papamastorakis J., A\&A, 344, 868
\bibitem[\protect\citeauthoryear{Wu \& Cao}{2006}]{2006PASP..118.1098W} Wu, Q., \& Cao, X. 2006, PASP, 118, 1098 
\bibitem[\protect\citeauthoryear{Zhu et al.}{2008}]{2008ApJ...686..155Z} Zhu, Y.-N., Wu, H., Cao, C., \& Li, H.-N. 2008, ApJ, 686, 155
\end{thebibliography}
\end{document}